\documentclass[11pt,twoside]{article}

\usepackage{asp2006}
\usepackage{epsf}
\usepackage{psfig}
\usepackage{lscape}

\begin{document}
\title{Gravitational-wave Astronomy: Opening a New Window on the Universe for
Students, Educators and the Public}
\author{M.\ Cavagli\`a$^1$, M.\ Hendry$^2$, D.\ Ingram$^3$, S.\ Milde$^4$, D.\
Reitze$^5$, K.\ Riles$^6$, B.\ Schutz$^7$, A.L.\ Stuver$^8$, T.\ Summerscales$^9$,
J.\ Thacker$^8$, C.V.\ Torres$^{8,10}$, D.\ Ugolini$^{11}$, M.\ Vallisneri$^{12}$,
A.\ Zermeno$^{13}$} 
\affil{$^1$ University of Mississippi, University, MS 38677, USA\\ $^2$
University of Glasgow, Glasgow, G12 8QQ, United Kingdom\\  $^3$ LIGO Hanford
Observatory, Richland, WA 99352, USA\\  $^4$ Milde Marketing Science
Communication, Hannover, Germany\\ $^5$ University of Florida, Gainesville, FL
32611, USA\\ $^6$ University of Michigan, Ann Arbor, MI 48109, USA\\ $^7$
Albert-Einstein-Institut, D-14476 Golm, Germany\\ $^8$ LIGO Livingston
Observatory, Livingston, LA 70754, USA\\ $^9$ Andrews University, Berrien
Springs, MI 49104 USA\\ $^{10}$ Louisiana State University, Baton Rouge, LA 70803, USA \\
$^{11}$ Trinity University, San Antonio, TX 78212, USA\\
$^{12}$ Jet Propulsion Laboratory, Pasadena, CA 91109, USA\\ $^{13}$ The
University of Texas at Brownsville, Brownsville, TX 78520, USA} 

\begin{abstract}
The nascent field of gravitational-wave astronomy offers many opportunities for effective and inspirational
astronomy outreach. Gravitational waves, the ``ripples in space-time'' predicted by Einstein's theory of
General Relativity, are produced by some of the most energetic and dramatic phenomena in the cosmos,
including black holes, neutron stars and supernovae.  The detection of gravitational waves will help to
address a number of fundamental questions in physics, from the evolution of stars and galaxies to the origin
of dark energy and the nature of space-time itself.  Moreover, the cutting-edge technology developed to
search for gravitational waves is pushing back the frontiers of many fields, from lasers and materials
science to high performance computing, and thus provides a powerful showcase for the attractions and
challenges of a career in science and engineering. For several years a worldwide network of ground-based
laser interferometric gravitational-wave detectors has been fully operational, including the two LIGO
detectors in the United States.  These detectors are already among the most sensitive scientific instruments
on the planet and in the next few years their sensitivity will achieve further significant improvement.
Those developments promise to open an exciting new window on the Universe, heralding the arrival of
gravitational-wave astronomy as a revolutionary, new observational field. In this paper we describe the
extensive program of public outreach activities already undertaken by the LIGO Scientific Collaboration, and
a number of special events which we are planning for IYA2009. 
\end{abstract}



\section{Gravitational Waves and LIGO}

Albert Einstein first described  gravitational waves in 1916 as part of his theory of General Relativity.
Gravitational waves are commonly referred to as ``ripples in the fabric of space-time''. They can be thought as
fluctuating distortions of the space-time geometry that are expected to propagate with the speed of light, and
can be visualized like waves on the surface of a pond.  Direct measurement of gravitational waves is currently
being attempted with a worldwide network of ground-based detectors, including the two LIGO (\emph{Laser
Interferometer Gravitational-wave Observatory}) detectors in the United States as well as the German-UK GEO600
detector in Germany and the French-Italian Virgo detector in Italy.

The LIGO Observatories in Hanford, WA and Livingston, LA attempt to measure the stretching of space-time by
means of a controlled laser beam. Their basic design is that of a Fabry-Perot Michelson interferometer: two
long orthogonal  ``arms'' forming an L shape, suspended mirrors at both arm ends and a beam splitter at
their common corner. Laser light enters the arms at the beam splitter and bounces between the mirrors
repeatedly before returning to the entry point \citep{Raab:2004}.  A gravitational wave impinging on the
interferometer causes the length of the two arms to vary in a predictable way, allowing detection through
the change in the interference pattern of the two beams. 

As a frontier physics effort, LIGO offers a unique opportunity to inspire interest in science among students
and to educate the broader community. The LIGO Scientific Collaboration (LSC) and the LIGO Observatories
support a broad program of education and outreach to take advantage of these opportunities. Planned activities
for the IYA2009 include programs at the Livingston Observatory  Science Education Center and at the Hanford
Observatory; programs on gravitational-wave astronomy for the classroom, interdisciplinary events linking
gravitational-wave astronomy to music and the visual arts; and research experiences for schools and citizens
through the ``Einstein@Home'' program.

\section{Programs at the LIGO Observatories}

The LIGO Observatories use the excitement of the search for gravitational waves as a platform for increasing
science interest and science literacy among all ages.  The grand scale of the LIGO detectors, the technology
necessary for their operation, and the innate public curiosity about black holes, supernovae and other
astrophysical exotica combine to make the Observatories a destination of interest for several thousand
visitors each year. Many of these public events and activities are undertaken in collaboration with partner
organizations. 

\noindent
{\bf General public.}  The Observatories offer regularly scheduled tours, star parties, periodic open houses
(including bilingual Spanish-English open houses at the Hanford Observatory) for the public. Both the Hanford
and Livingston Observatories participate in off-site venues such as community festivals, service club meetings
and other municipal sponsored venues.

\noindent
{\bf Grades K-12.} 
Field trips to the Observatories include a walking tour of the facility, a visit to the control room,
interactions with LIGO personnel and hands-on experiences with a number of exhibits.  The exhibits match the
themes of LIGO science and are correlated to national science education standards. The Observatories also offer
classroom visits during which students can explore the physics of waves through additional hands-on activities.
LIGO is a partner in the NSF-funded I2U2 program that makes environmental data available for student research
projects.

\noindent
{\bf Undergraduate.} The Hanford and Livingston Observatories host visits from physics undergraduate groups
whose institutions range from Washington to Montana, and from Louisiana to Mississippi, respectively.  The
Livingston Observatory offers undergraduates from Southern University opportunities to assist with its
outreach activities by serving as docents.  Undergraduate internships are available that provide intensive
summer research experience.  

\noindent
{\bf Educators.}  Professional development activities that emphasize the nature of scientific inquiry form
the basis of LIGO's involvement with teachers.  The Observatories partner with regional higher education
institutions, local school districts and school support providers for the delivery of these programs. 
Summer scientific research programs are also available to educators.

\section{Programs on Gravitational-wave Astronomy for the Classroom}

There are three main experiential learning classroom activities that are used to illustrate interferometric
gravitational-wave detector instrumentation and data analysis. Links to these are currently hosted on the
Einstein's Messengers website (http://www.einsteinsmessengers.org/activities.htm). Einstein's Messengers is an
educational video documentary, produced by the NSF, on LIGO and the potential of gravitational-wave astronomy. 
For each activity there are references in the teacher instructions to specific times in the documentary that are
relevant to it.

\noindent
{\bf A model Michelson interferometer.} This activity puts the student in the place of the
gravitational-wave experimentalist by giving students hands-on experience in building a small scale
Michelson interferometer to investigate interference patterns and basic LIGO operation. Students learn
interference concepts, identify how mirror motions affect the fringe patters, and become familiar with the
concept of strain and how LIGO will detect gravitational waves.

\noindent
{\bf Searching for gravitational waves in noisy data.} This activity puts the student in the place of a LIGO
data analyst by introducing students to compact binary coalescence sources and the matched filtering methods
used to seek these signals in the predominant detector noise \citep{Larson:2006}. Students learn the concept
of signal and noise in scientific measurements, data filtering to determine the likelihood of detecting a
signal and how LIGO scientists would use these methods to search for gravitational waves.

\noindent
{\bf Extracting astrophysical information from simulated gravi\-ta\-tio\-nal-wave signals.} This activity
puts the student in the place of a gravita\-tio\-nal-wave astronomer by having students use scientific
theory to extract information on the physical system that produced a detected gravitational wave.  Students
learn how gravitational-wave signals from compact binary systems evolve in time, use simulated signal
measurements and relevant equations to determine the chirp masses and astronomical distances of the systems,
and experience the scientific process that will be the underpinnings of gravitational-wave astronomy.

All of these activities have a target audience of secondary education students, but can easily be adapted for
students in the mid- to upper-primary education levels.  Students can then share with their colleagues their
findings and explain their rationale thus engaging them as real scientists.  Such an adaptation has been
successful with students as young as 4th grade.

\section{Interdisciplinary Events: Blending Art and Science}

Exhibits and public events blending art and science help to increase the understanding of gravitational-wave
astronomy in an unconventional way.

In partnership with professional designers, the LSC is devising an exhibit that will connect onlookers and
passersby to the science central to LIGO research through a creative lighting art work. This exhibit will first
be displayed at a central location in New York City during the 2009 World Science Festival
(http://www.worldsciencefestival.org) and later permanently relocated to a public institution or science museum.
A signature event featuring gravitational waves, produced by the World Science Festival, will complement the
exhibit. 

Andrea Centazzo (http://www.andreacentazzo.com), an internationally re\-no\-wned composer and multimedia artist, is
developing a show about General Relativity and gravitational waves. The show involves Andrea playing live in
sync with a video montage of ``real world" footage and animations. The LSC is planning to pair it with public
lectures by a LIGO scientist to bring together scientifically-oriented and artistically-oriented audiences. The
show premi\`ere will be at Caltech's Beckman Auditorium in Fall 2008. 

\section{Einstein@Home: Anyone Can Search for Gravitational Waves}

Einstein@Home (http://einstein.phys.uwm.edu) uses the BOINC (http://boinc. berkeley.edu) distri\-bu\-ted
computing software to utilize computer time donated by computer users all over the world to process data from
gravitational-wave detectors. Participants in Einstein@Home download the BOINC software to their computers and
join the Einstein@Home project, which processes gravitational-wave data when the computer is not being used for
other applications, like word processors or games. Einstein@Home does not affect the performance of computers
and greatly speeds up this exciting research and has already published results \citep{LSC:2008}.

\acknowledgements

LIGO was constructed by the California Institute of Technology and Massachusetts Institute of Technology with
funding from the National Science Foundation and operates under cooperative agreement PHY-0107417. The authors
gratefully acknowledge the support of the National Science Foundation through LIGO Research Support (LSC). 
This paper has LIGO Document Number LIGO-P080059-00-Z.


\end{document}